\documentclass[12pt]{article}
\pdfoutput=1
\usepackage{jcappub}
\usepackage{amsmath}
\usepackage{amssymb}
\usepackage{physics}
\usepackage{xcolor}
\usepackage{slashed}
\usepackage{subfigure}
\usepackage{float}
\usepackage{notoccite}

\definecolor{dgreen}{HTML}{008000}

\def\be{\begin{equation}}
\def\ee{\end{equation}}
\def\bea{\begin{eqnarray}}
\def\eea{\end{eqnarray}}

\title{Bargmann Representation of Spin Chains}

\author[1]{M. W.  AlMasri }
\emailAdd{mwalmasri2003@gmail.com}\author[1,2]{and M. R. B.  Wahiddin} 

\affiliation[1]{Cybersecurity and  Systems Research Unit, ISI-USIM , Bandar Baru Nilai, 71800 Nilai, Negeri
	Sembilan, Malaysia}
\affiliation[2]{Pusat Tamhidi, USIM, Bandar Baru Nilai, 71800 Nilai, Negeri
	Sembilan, Malaysia }

\begin{document}

\abstract{
	Spin chain Hamiltonians can be written in terms of complex differential operators using the Bargmann representation of the Jordan-Schwinger map. In this case, the eigenfunctions are expressed as the product of orthonormal monomials of the phase-space coordinates $z_{i}=Q_{i}+i P_{i}$ in the complex plane. Furthermore, the series constructed from each phase-space coordinate converges uniformly in any compact domain of the complex plane. Formulating spin chains with respect to the phase-space coordinates helps in discussing their classical limit and in the calculations of quasi-probability distributions.}
\maketitle
\section{Introduction}
Phase-space formulation of quantum theory appears to be very effective in studying some aspects related to probability distributions in quantum optics \cite{Zachos}.  Furthermore, the general shape of quantum mechanics formalism becomes similar to  classical mechanics when we formulate it using complex canonical coordinates. This formulation is very suitable for discussing the classical limit of quantum theory \cite{Strocchi}. 
One related interesting formulation of quantum mechanics is  the  holomorphic representation  introduced  by Bargmann and Segal  in 60s\cite{Bargmann,Segal,Hall}. The main idea behind this construction is to represent  wave functions as monomials in the complex plane i.e. $\psi_{n}(z)= \frac{z^{n}}{\sqrt{n!}}$ not as matrices. Here $z$ is a complex phase-space variable i.e.   $z=\hat{Q}+i\hat{P}$, where   $\hat{Q}$ and $\hat{P}$ are the  configuration  and momentum space operators respectively.   In this case, the raising and lowering operators are $a^{\dagger}=z$ and $a=\frac{\partial}{\partial z}$ respectively\footnote{ Since for any given  arbitrary complex  function $f(z)$ we have
\begin{equation}
	\left[\frac{\partial }{\partial z}, z\right]f= \frac{\partial }{\partial z}\left(z f\right)- z \frac{\partial f}{\partial z}= f
\end{equation}} 
. Thus $[\frac{\partial }{\partial z}, z]=1$ has the same structure similar to the commutator of $a$ and $a^{\dagger}$. This assignment of raising and lowering operators  was originally proposed by Fock \cite{Fock}. 
Moreover, the series  $\psi_{n}(z)=\sum c_{n}\frac{z^{n}}{\sqrt{n!}}$ converges uniformly   in any compact domain of the complex $z$-plane  since $\sum^{\infty}_{n=0} |c_{n}|^{2}=1$ \cite{Boas,Askold,Folland}. Possible applications of  Bargmann representation in various physical systems can be found in \cite{Stenholm,Bishop,Voros,Thiemann,Kowalski, almasri}.\vskip 5mm

\section{Bargmann Representation}

{\bf Definition:}  {\it The Bargmann  or  Segal-Bargmann spaces   $\mathcal{H}L^{2}(\mathbb{C}^{d},\mu)$ are   spaces of holomorphic functions with Gaussian integration measure $\mu= (\pi)^{-d} e^{-|z|^{2}}$  such that the inner-product endowed with this space is } \cite{Bargmann,Segal,Hall,Askold}
\begin{equation}\label{inner}
	\langle f|g\rangle_{\mu}= (\pi)^{-d}\int_{\mathbb{C}^{d}} \overline{f}(z)\;  g(z) e^{-|z|^{2} } dz, 
\end{equation}
where $|z|^{2}=|z_{1}|^{2}+ \dots+ |z_{d}|^{2}$.\vskip 5mm
Any entire  function $f(z)$ in $\mathcal{H}L^{2}(\mathbb{C}^{d},\mu)$ obeys   the following  square-integrability condition
\begin{equation}\label{condition}
	||f||^{2}: =\langle f| f\rangle_{\mu}= (\pi)^{-d}\int_{\mathbb{C}^{d}} |f(z)|^{2} e^{-|z|^{2}}dz < \infty,
\end{equation}
where  $dz$ is the $2d$-dimensional Lebesgue measure in  $\mathbb{C}^{d}$. The Bargmann space $\mathcal{H}L^{2}(\mathbb{C}^{d}, \mu)$ is in fact  a Hilbert space as shown by Bargmann in \cite{Bargmann}. Using the inner-product defined in \ref{inner}, we can prove that both  $\overline{z}$ and $\frac{\partial}{\partial z}$ have the same effect i.e. $\langle\frac{\partial f}{\partial z}, g\rangle_{\mu} = \langle f,z g\rangle_{\mu}$ \cite{Hall}. We define 
the inverse Segal-Bargmann transform as a unitary  map $B^{-1}: \mathcal{H}L^{2}(\mathbb{C}^{d},\mu)\rightarrow L^{2}(\mathbb{R}^{d},dx)$ \cite{Hall}
\begin{equation}
	B^{-1}f(x)= \int_{\mathbb{C}}\exp[-(\overline{z}\cdot\overline{z}-2\sqrt{2}\overline{z}\cdot x+x\cdot x)/2] f(z)  e^{-|z|^{2}}\; dz.
\end{equation}

\vskip 5mm
The harmonic oscillator Hamiltonian  $\hat{H}=\hbar \omega \left(\hat{a}^{\dagger}\hat{a}+ \frac{1}{2}\right)$ assumes the following form in the Bargmann representation 
\begin{equation}\label{harmonic}
	\hat{H}= \hbar \omega \left(z \frac{d}{dz}+ \frac{1}{2}\right)
\end{equation}
In this case, the orthonormal eigenfunctions are 
 $\{z^{n}/\sqrt{n!}\}$  and the orthonormality relation is 
\begin{equation}
	\int_{\mathbb{C}}  e^{-|z|^{2}} \; \overline{z}^{n} z^{m} dz = n! \pi\; \delta_{mn}. 
\end{equation}

and the corresponding energy eigenvalues are 
\begin{equation}
	\hat{H}|n\rangle= \hbar \omega \left(z \frac{d}{dz}+ \frac{1}{2}\right) \frac{z^{n}}{\sqrt{n!}} = \hbar \omega \left(n+\frac{1}{2}\right)\frac{z^{n}}{\sqrt{n!}}= \hbar \omega \left(n+\frac{1}{2}\right) |n\rangle.
\end{equation}
The  ground-state wave function in the  Bargmann  representation is simply $\psi_{0}(z)= 1$  while in coordinate representation it is $\psi_{0}(x)= \left(\frac{m\omega }{2\hbar}\right)^{1/4} e^{-m \omega x^{2}/2\hbar}$. Generally ,  the wave functions of quantum  harmonic oscillator in Bargmann  representation $\psi_{n}(z)=\frac{z^{n}}{\sqrt{n!}}$ correspond to   $\psi_{n}(x)= \frac{1}{\sqrt{2^{n} n!}} \left(\frac{m\omega}{\pi \hbar}\right)^{\frac{1}{4}} e^{-\frac{m\omega x^{2}}{2\hbar}} H_{n}(\sqrt{\frac{m \omega }{\hbar}}x) $ where $H_{n}(y)= (-1)^{n} e^{y^{2}} \frac{d^{n}}{dy^{n}} (e^{-y^{2}})$ is the Hermite polynomials in coordinate representation. \vskip 5mm
Since $\{z^{n}\}$ forms an orthogonal basis, their magnitude carries no information. Therefore, only the exponents $n$ are physically important. Thus, we can impose conditions on $z_{i}$ so that we can safely make Taylor expansion of the  function $f(z_{i})$ valid and use it as energy eigenstates. 
Many famous functions in mathematics give series expansion in terms of monomials $z_{i}$ and their powers \cite{tables}. Thus we may extend  energy eigenfunctions  to include these as possible solutions. For example, consider  the series expansion of  the exponential function $e^{z}= \sum_{n=0}^{\infty} \frac{z^{n}}{n!}=\sum_{n=0}^{\infty}\frac{\psi(z)}{\sqrt{n!}}= \sum_{n=0}^{\infty } \phi(z)$  as possible energy eigenstates where $\psi(z)= \frac{z^{n}}{\sqrt{n!}}$ and $\phi(z)= \frac{z^{n}}{n!}$. Then  the expectation value $ \sum_{n=0}^{\infty } n!  \langle\phi(z)|\hbar \omega \left(z \frac{d}{dz}+ \frac{1}{2}\right) |\phi(z)\rangle_{\mu}$  gives   the summation of energy eigenvalues  of  the harmonic oscillator $E=(\frac{1}{2}+ \frac{3}{2}+ \frac{5}{2}+ \dots)$ as expected. Similarly, considering the series expansion of the functions  $\sinh(z)= \sum_{n=0}^{\infty}\frac{z^{2n+1}}{(2n+1)!}$ and $\cosh(z)= \sum_{n=0}^{\infty}\frac{z^{2n}}{(2n)!}$ give $E=(\frac{3}{2}+ \frac{7}{2}+ \frac{9}{2}+ \dots )$ and $E=(\frac{1}{2}+ \frac{5}{2}+ \frac{11}{2}+ \dots)$ respectively using proper normalization constants (namely multiplying the expectation value of the Hamiltonian by $2n!$ for $\cosh(z)$ and $(2n+1)!$ for $\sinh(z)$). One could consider other functions and combinations.  For example,  the series expansion of $\sin(z)= z- \frac{z^{3}}{3!}+ \frac{z^{5}}{5!}- \dots  $ or $\cos(z)= 1-\frac{z^{2}}{2!}+ \frac{z^{4}}{4!}- \dots $.  The expectation value of the Hamiltonian with a proper normalization  gives the same overall result as previous series expansion of  $\sinh(z)$ and $\cosh(z)$ respectively. The minus sign in front of some monomials can be absorbed automatically and has no effect on the energy eigenvalues, consider for example  the state $|\xi(z)\rangle = - z$, 
\begin{equation}
	\hat{H}|\xi\rangle= - \hbar \omega \left(z \frac{d}{dz}+ \frac{1}{2}\right) z= \frac{3}{2}\hbar \omega (-z)= \frac{3}{2}\hbar \omega |\xi\rangle  
\end{equation}
\section{Angular Momentum Operators in The Bargmann Space}
Consider a pair of uncoupled harmonic oscillators with annihilation operators $a$ and $b$ respectively. The commutation relations are 
\begin{eqnarray}\label{commutations}
	[a,a^{\dagger}]=[b,b^{\dagger}]=1, \; \; [a,a]=[a^{\dagger},b]=[a,b^{\dagger}]=[b,b]=0. 
\end{eqnarray}
We define the angular momentum operators as
\begin{eqnarray}
	J_{1}=\frac{\hbar }{2}\left(a^{\dagger}b+b^{\dagger}a\right), \\
	J_{2}= \frac{\hbar}{2i}\left(a^{\dagger}b-b^{\dagger}a\right),\\
	J_{3}=\frac{\hbar}{2}\left(a^{\dagger}a-b^{\dagger}b\right).
\end{eqnarray}
It can be shown in a straightforward manner taking into account \ref{commutations} that such construction obeys the canonical commutation relations for angular momentum 
\begin{eqnarray}
	[J_{i},J_{j}]=i\hbar\epsilon_{ijk}J_{k}, \; [J_{i},J^{2}]=0, 
\end{eqnarray}
where $i=1,2,3$ and $J^{2}$ is the squared  total angular momentum operator (also known as the Casimir element or invariant) . The total number operator is 
\begin{eqnarray}
	N=N_{a}+N_{b}=a^{\dagger}a+b^{\dagger}b, 
\end{eqnarray}
with integer eigenvalues $n,n_{a},n_{b}= 0,1,2,\dots$. In terms of number operators we may write 
\begin{eqnarray}
	J^{2}=\frac{\hbar^{2}N}{2}\left(\frac{N}{2}+1\right), \; J_{3}=\hbar\left(\frac{N_{a}-N_{b}}{2}\right)
\end{eqnarray}
Comparing previous relations with 
\begin{eqnarray}
	J_{3}|j,m\rangle=\hbar m|j,m\rangle,\\
	J^{2}|j,m\rangle=\hbar^{2} j(j+1)|j,m\rangle, 
\end{eqnarray}
we set  $j=\frac{n_{a}+n_{b}}{2}$ and $ m=\frac{n_{a}-n_{b}}{2}$. 
 \vskip 5mm
For a given entire  function $f_{\alpha,\beta}(z,w)= \frac{z^{\alpha}}{\sqrt{\alpha !}}\frac{w^{\beta}}{\sqrt{\beta!}}$,  we  define the angular momentum operators in the two-dimensional  Bargmann space $\mathcal{H}L^{2}(\mathbb{C}^{2}, \mu)$ by writing  the Jordan-Schwinger  map in the holomorphic representation  as \cite{Sakurai}
\begin{eqnarray}\label{1}
	\hat{J}_{1}= \frac{\hbar}{2} \left(z \frac{\partial }{\partial w}+ w \frac{\partial}{\partial z}\right), \\\label{2}
	\hat{J}_{2}= \frac{\hbar}{2i} \left(z \frac{\partial }{\partial w}-w \frac{\partial }{\partial z}\right), \\\label{3}
	\hat{J}_{3}= \frac{\hbar}{2}\left(z \frac{\partial }{\partial z}- w\frac{\partial}{\partial w}\right). 
\end{eqnarray}
These operators belong to the $SU(2)$ Lie algebra and obey the following commutation relations $[\hat{J}_{i},\hat{J}_{j}]= i\hbar\; \varepsilon_{ijk }\hat{J}_{k}$ since the only non-trivial commutators between $z$, $w$ and their partial derivatives  are $\left[\frac{\partial}{\partial z},z\right]=\left[\frac{\partial }{\partial w},w\right]=1$ . $\varepsilon_{ijk }$ is the well-known Levi-Civita totally antisymmetric symbol. 
With this construction, we can  express the non-Hermitian  raising and lowering operators as $\hat{J}_{+}= \hat{J}_{1}+ i \hat{J}_{2}= \hbar z \frac{\partial }{\partial w}$ and $\hat{J}_{-}= \hat{J}_{1}-i\hat{J}_{2}= \hbar w \frac{\partial }{\partial z}$ .

\vskip 5mm
The inner product in the two-dimensional Bargmann space $\mathcal{H}L^{2}(\mathbb{C}^{2}, \mu)$ is 
\begin{align}\label{ortho}
	\langle f_{\alpha^{\prime},\beta^{\prime}}(z,w)|f_{\alpha,\beta}(z,w)\rangle_{\mu}\\ \nonumber=& \frac{1}{\pi^{2}}\int dz \; dw \exp[- |z|^{2} -|w|^{2}] \overline{f_{\alpha^{\prime}, \beta^{\prime}}(z,w)} f_{\alpha,\beta}(z,w) =   \delta_{\alpha^{\prime},\alpha}\delta_{\beta^{\prime},\beta}
\end{align}
Let us explicitly calculate  the expectation values of the operators  $\hat{J}_{3}$(or $\hat{J}_{z}$ as called in most of references), $\hat{J}^{2}$ and $\hat{J}_{\pm}$  using the basis $f_{\alpha,\beta}(z,w)$, we find 
\begin{align}
	\langle f_{\alpha^{\prime},\beta^{\prime}}| \hat{J}_{3}|f_{\alpha,\beta}\rangle_{\mu}= \frac{\hbar }{2}\left(\alpha -\beta \right)\; \delta_{\alpha^{\prime},\alpha} \delta_{\beta^{\prime},\beta}, \\
	\langle f_{\alpha^{\prime},\beta^{\prime}}| \hat{J}^{2}|f_{\alpha,\beta}\rangle_{\mu}= \hbar^{2} \left(\frac{\alpha+\beta}{2}\right)\left(\frac{\alpha+\beta}{2}+1\right)\; \delta_{\alpha^{\prime},\alpha} \delta_{\beta^{\prime},\beta},\\ 
	\langle f_{\alpha^{\prime},\beta^{\prime}}| \hat{J}_{+}|f_{\alpha,\beta} \rangle_{\mu}= \hbar \sqrt{(\alpha+1)\beta}\; \delta_{\alpha^{\prime},\alpha+1} \delta_{\beta^{\prime},\beta-1},\\
	\langle f_{\alpha^{\prime},\beta^{\prime}}| \hat{J}_{-}|f_{\alpha,\beta} \rangle_{\mu}= \hbar \sqrt{\alpha (\beta+1)}\;  \delta_{\alpha^{\prime}, \alpha-1}\delta_{\beta^{\prime}, \beta+1}.
\end{align} 
In contrast to  the standard treatment of angular momentum mentioned for example in \cite{Sakurai}, we can directly compute the expectation values of  operators $\hat{J}_{1,2}$ without invoking $\hat{J}_{\pm}$ and $\hat{J}_{3}$ in our calculations. We obtain 

\begin{align}
	\langle f_{\alpha^{\prime},\beta^{\prime}}| \hat{J}_{1}|f_{\alpha,\beta} \rangle_{\mu}  = \frac{\hbar}{2}(\sqrt{(\alpha+1)\beta}\; \delta_{\alpha^{\prime},\alpha+1}\; \delta_{\beta^{\prime},\beta-1} + \sqrt{\alpha (\beta+1)}\; \delta_{\alpha^{\prime},\alpha-1}\delta_{\beta^{\prime},\beta+1}), \\
	\langle f_{\alpha^{\prime},\beta^{\prime}}| \hat{J}_{2}|f_{\alpha,\beta} \rangle_{\mu}= \frac{\hbar }{2i} (\sqrt{(\alpha+1)\beta}\; \delta_{\alpha^{\prime},\alpha+1}\delta_{\beta^{\prime},\beta-1}- \sqrt{\alpha(\beta+1)} \; \delta_{\alpha^{\prime},\alpha-1}\delta_{\beta^{\prime},\beta+1}).
\end{align}
Comparing $\hat{J}_{3}$ and $\hat{J}^{2}$ with the standard relations written in  $|jm\rangle$ basis  i.e. $\hat{J}_{z}|jm\rangle= m\hbar |jm\rangle$ and $\hat{J}^{2}|jm\rangle= \hbar^{2} j(j+1)|jm\rangle$,  we find 
\begin{eqnarray}\label{exponent}
	j= \frac{\alpha+\beta}{2},\;\; 
	m= \frac{\alpha-\beta}{2}.
\end{eqnarray}
For spin-$1/2$ particles we have $j=1/2$, this implies either $\alpha=1,\beta=0$ which corresponds to $m=1/2$ and $f_{10}= z$ or $\alpha=0,\beta=1$ which gives $m=-1/2$ and $f_{01}=w$. Analogously for $j=1$ we have three different case I) $\alpha=1,\beta=1$ which corresponds to $m=0$ and $f_{11}=zw$, II) $\alpha=2$,$\beta=0$ with $m=1$ and $f_{20}=\frac{z^{2}}{\sqrt{2}}$ , finally III) when $\alpha=0,\beta=2$ and this gives $m=-1, f_{02}=\frac{w^{2}}{\sqrt{2}}$. The computation for $j=N$ is straightforward, one should take into account that we have $2j+1$ number of possible states. For $j=N$, where $N$ is a positive integer (Bosons), $m=0$ is equivalent to $\alpha=\beta$ and in this case the energy eigenfunctions can be written as $\frac{(z w)^{n}}{n!}$. Note that the series expansion of $e^{zw}=\sum_{j=0}^{\infty} \frac{(z w)^{n}}{n!}$  gives all the  eigenfunctions with $m=0$ and  $j$ running  from 0 to $\infty$. 
\vskip 5mm
The  addition of quantum angular momentum  using   holomorphic representation can be done in a systematic way by adding monomials without  worrying about the dimensionality  of tensor product of single-particle angular momentum subspaces. For simplicity we consider a two-particle system. The total angular momentum operator is 
\begin{equation}
	\hat{J}= \hat{J}_{1}+ \hat{J}_{2}
\end{equation}
where 1 and 2 labels the two particles. we assume the joint two-particle system basis to be $|f_{\alpha_{1},\alpha_{2},\beta_{1},\beta_{2}}\rangle= \frac{z^{\alpha_{1}}_{1}}{\sqrt{\alpha_{1}!}}\frac{z^{\alpha_{2}}_{2}}{\sqrt{\alpha_{2}!}}\frac{w^{\beta_{1}}_{1}}{\sqrt{\beta_{1}!}}\frac{w^{\beta_{2}}_{2}}{\sqrt{\beta_{2}!}}$, then the operators  $\hat{J}^{2}$ and $\hat{J}_{3}$ are 

\begin{align}\label{j2}
	\hat{J}^{2}|f_{\alpha_{1},\alpha_{2},\beta_{1},\beta_{2}}\rangle= \left(\hat{J}^{2}_{1}+\hat{J}^{2}_{2}\right)|f_{\alpha_{1},\alpha_{2},\beta_{1},\beta_{2}}\rangle = \hat{J}^{2}_{1}|f_{\alpha_{1}\beta_{1}} \rangle+ \hat{J}^{2}_{2}|f_{\alpha_{2}\beta_{2}}\rangle\\ \nonumber= \hbar^{2} \left(\frac{\alpha_{1}+\beta_{1}}{2}\right)\left(\frac{\alpha_{1}+\beta_{1}}{2}+1\right) |f_{\alpha_{1}\beta_{1}} \rangle+  \hbar^{2} \left(\frac{\alpha_{2}+\beta_{2}}{2}\right)\left(\frac{\alpha_{2}+\beta_{2}}{2}+1\right) |f_{\alpha_{2}\beta_{2}}\rangle\\
	\nonumber = \hbar^{2} \left(\frac{{\bf \alpha}+{\bf \beta}}{2}\right)\left(\frac{{\bf \alpha}+{\bf \beta}}{2}+1\right)|f_{\alpha_{1},\alpha_{2},\beta_{1},\beta_{2}}\rangle, 
\end{align}

\begin{eqnarray}\label{j3}
	\hat{J}_{3}|f_{\alpha_{1},\alpha_{2},\beta_{1},\beta_{2}}\rangle= \hbar \frac{\left(\alpha-\beta\right)}{2}|f_{\alpha_{1},\alpha_{2},\beta_{1},\beta_{2}}\rangle, 
\end{eqnarray}
where $\alpha=\alpha_{1}+\alpha_{2}$ and $\beta=\beta_{1}+\beta_{2}$. \vskip 5mm The generalization to $N$-particle system is very natural in the analytical approach. The previous relations \ref{j2} and \ref{j3} generalize to 
\begin{align}
	\hat{J}^{2}|f_{\alpha_{1}\dots \alpha_{N},\beta_{1}\dots \beta_{N}}\rangle=\hbar^{2} \left(\frac{{\bf \alpha}+{\bf \beta}}{2}\right)\left(\frac{{\bf \alpha}+{\bf \beta}}{2}+1\right)|f_{\alpha_{1}\dots \alpha_{N},\beta_{1}\dots \beta_{N}}\rangle, \\ 
	\hat{J}_{3}|f_{\alpha_{1}\dots \alpha_{N},\beta_{1}\dots \beta_{N}}\rangle= \hbar \frac{\left(\alpha-\beta\right)}{2} |f_{\alpha_{1}\dots \alpha_{N},\beta_{1}\dots \beta_{N}}\rangle,
\end{align}
where here $\alpha=\alpha_{1}+\dots+ \alpha_{N}$ and $\beta=\beta_{1}+ \dots + \beta_{N}$. 
\vskip 5mm
Having established the angular momentum operators and their main addition relations in the holomorphic representation,  it is a straightforward procedure to reproduce all known formulas  related to quantum angular momentum such as the Clebsch-Gordon coefficients, Wigner 3-$j$ , 6-$j$ and Racah symbols by means of analytical functions and their partial  derivatives\cite{Biedenharn}. Furthermore using the holomorphic representation of Jordan-Schwinger mapping  we may find the  holomorphic representation of spin chains  \cite{Lieb1,Nepomechie}and topologically protected magnetic solitons such as magnetic skyrmions \cite{Nagaosa,thesis, skyrmion}.  

\section{Spin Chains }
Now, consider the general  $XYZ$-spin chain Hamiltonian 
\begin{eqnarray}\label{xyz}
	\hat{H}_{XYZ}= \sum_{i} \left(J_{x} \hat{S}^{x}_{i}\hat{S}^{x}_{i+1}+ J_{y} \hat{S}^{y}_{i}\hat{S}^{y}_{i+1}+ J_{z} \hat{S}_{i}^{z} \hat{S}_{i+1}^{z}\right), 
\end{eqnarray}
where $i$ is the site index. Applying the transformations \ref{1},\ref{2},\ref{3} into \ref{xyz} give the holomorphic representation of the $XYZ$-spin chain ,  
\begin{align}\label{xyz1}
	\hat{H}_{XYZ}= \frac{\hbar^{2}J_{x}}{4}\sum_{i} \left(z_{i}z_{i+1} \frac{\partial^{2}}{\partial w_{i}\partial w_{i+1}}+ w_{i}z_{i+1}\frac{\partial^{2}}{\partial z_{i}\partial w_{i+1}}+ w_{i+1}z_{i} \frac{\partial^{2}}{\partial z_{i+1}\partial w_{i}}+w_{i}w_{i+1}\frac{\partial^{2}}{\partial z_{i}\partial z_{i+1}}\right)\\ \nonumber
	-\frac{\hbar^{2}J_{y}}{4} \sum_{i}\left(z_{i}z_{i+1}\frac{\partial^{2}}{\partial w_{i}\partial w_{i+1}}- w_{i}z_{i+1}\frac{\partial^{2}}{\partial z_{i}\partial w_{i+1}}-z_{i}w_{i+1} \frac{\partial^{2}}{\partial w_{i}\partial z_{i+1}}+ w_{i}w_{i+1}\frac{\partial^{2}}{\partial z_{i}\partial z_{i+1}}\right)\\ \nonumber 
	+ \frac{\hbar^{2}J_{z}}{4} \sum_{i}\left(z_{i}z_{i+1}\frac{\partial^{2}}{\partial z_{i}\partial z_{i+1}}-2 w_{i} z_{i+1} \frac{\partial^{2}}{\partial w_{i}\partial z_{i+1}}+ w_{i}w_{i+1}\frac{\partial^{2}}{\partial w_{i}\partial w_{i+1}}\right)
\end{align}

When $J_{x}=J_{y}=J_{z}$, \ref{xyz1} reduces to the $XXX$ Spin chain Hamiltonian , 
\begin{eqnarray}
	\hat{H}_{XXX}= \frac{\hbar^{2} J_{x}}{4}\left(z_{i}z_{i+1}\frac{\partial^{2}}{\partial z_{i}\partial z_{i+1}}+ 2z_{i}w_{i+1}\frac{\partial^{2}}{\partial w_{i}\partial z_{i+1}}+ w_{i}w_{i+1}\frac{\partial^{2}}{\partial w_{i}\partial w_{i+1}}\right)
\end{eqnarray} 
with eigenfunctions written as 
\begin{eqnarray}
	f^{i}_{\alpha\beta\gamma\delta}=\frac{z^{\alpha}_{i} z_{i+1}^{\beta}w_{i}^{\gamma}w_{i+1}^{\delta}}{\sqrt{\alpha! \beta! \gamma! \delta!}}. 
\end{eqnarray}

The expectation value of $XXX$ spin  chain Hamiltonian is 
\begin{align}
	\langle f^{i}_{\alpha^{\prime}\beta^{\prime}\gamma^{\prime}\delta^{\prime}}| \hat{H}_{XXX}|f^{j}_{\alpha\beta\gamma\delta}\rangle_{\mu}=\frac{\hbar^{2}J_{x}\left(\alpha+\beta\right)}{4} \sum_{i}  \langle f^{i}_{\alpha^{\prime}\beta^{\prime}\gamma^{\prime}\delta^{\prime
	}}| f^{j}_{\alpha\beta\gamma\delta}\rangle_{\mu}\delta_{ij}+\\ \nonumber \frac{\hbar^{2}J_{x}\gamma\delta}{2} \sum_{i} \langle f^{i}_{\alpha^{\prime}\beta^{\prime}\gamma^{\prime}\delta^{\prime
	}}|f^{j}_{\alpha+1 \beta-1 \gamma+1 \delta-1}\rangle_{\mu}\delta_{ij}
\end{align}
where $\langle f^{i}_{\alpha^{\prime}\dots}|f^{j}_{\kappa\dots}\rangle_{\mu}=\delta_{ij}\delta_{\alpha\kappa\dots}$. From the holomorphic representation of $\hat{H}$, we can compute the partition function from which we may determine thermodynamics quantities such as entropy and free energy... .  \section{Conclusion} In  this note, we presented a general procedure for formulating spin chains  in the space of holomorphic functions with Gaussian integration measure know as Bargmann representation. 
 One advantage of the Bargmann  representation is the analog of quantum and classical formalism as emphasized in \cite{Strocchi}. Another privilege of the Bargmann representation stems from the fact that all operators are written in terms of the phase-space canonical  coordinates. Since quasi-probability distributions such as the Glauber–Sudarshan $P$-representation
and Husimi-Kano $Q$-representations are written in the phase-space. Thus, Bargmann representation is the natural home for studying these quantities which has deep applications in quantum optics .  \vskip 5mm 
{\bf Acknowledgment:}  We are grateful to USIM for financial support.

\end{document}